\documentclass[sigconf]{acmart}

\usepackage[T1]{fontenc}
\usepackage{enumitem}
\usepackage{afterpage}
\usepackage{listings}
\usepackage{xcolor}
\usepackage{graphicx}
\usepackage{float}
\usepackage[font=footnotesize]{caption}
\usepackage[font=footnotesize]{subcaption}
\usepackage{placeins}
\usepackage{url}
\usepackage{epstopdf}
\usepackage{booktabs}
\usepackage{footnote}
\makesavenoteenv{table}
\makesavenoteenv{tabular}

\usepackage[acronyms,nonumberlist,nopostdot,nomain,nogroupskip]{glossaries}

\usepackage{tabularx}

\usepackage{paralist}

\newacronym{3gpp}{3GPP}{3rd Generation Partnership Project}
\newacronym{adc}{ADC}{Analog to Digital Converter}
\newacronym{5g}{5G}{5th generation}
\newacronym{aimd}{AIMD}{Additive Increase Multiplicative Decrease}
\newacronym{am}{AM}{Acknowledged Mode}
\newacronym{amc}{AMC}{Adaptive Modulation and Coding}
\newacronym{aqm}{AQM}{Active Queue Management}
\newacronym{awgn}{AGWN}{Additive White Gaussian Noise}
\newacronym{balia}{BALIA}{Balanced Link Adaptation}
\newacronym{bdp}{BDP}{Bandwidth-Delay Product}
\newacronym{bf}{BF}{Beamforming}
\newacronym{cc}{CC}{Congestion Control}
\newacronym{cdf}{CDF}{Cumulative Distribution Function}
\newacronym{cn}{CN}{Core Network}
\newacronym{cqi}{CQI}{Channel Quality Information}
\newacronym{cp}{CP}{Control Plane}
\newacronym{csirs}{CSI-RS}{Channel State Information - Reference Signal}
\newacronym{dc}{DC}{Dual Connectivity}
\newacronym{dce}{DCE}{Direct Code Execution}
\newacronym{dci}{DCI}{Downlink Control Information}
\newacronym{dl}{DL}{Downlink}
\newacronym{dmr}{DMR}{Deadline Miss Ratio}
\newacronym{dmrs}{DMRS}{DeModulation Reference Signal}
\newacronym{e2e}{E2E}{End-to-End}
\newacronym{ecn}{ECN}{Explicit Congestion Notification}
\newacronym{edf}{EDF}{Earliest Deadline First}
\newacronym{enb}{eNB}{evolved Node Base}
\newacronym{epc}{EPC}{Evolved Packet Core}
\newacronym{es}{ES}{Edge Server}
\newacronym{fdma}{FDMA}{Frequency Division Multiple Access}
\newacronym{fdd}{FDD}{Frequency Division Duplexing}
\newacronym[firstplural=Radio Access Technologies (RATs)]{rat}{RAT}{Radio Access Technology}
\newacronym{fs}{FS}{Fast Switching}
\newacronym{ftp}{FTP}{File Transfer Protocol}
\newacronym{gnb}{gNB}{Next Generation Node Base}
\newacronym{harq}{HARQ}{Hybrid Automatic Repeat reQuest}
\newacronym{hetnet}{HetNet}{Heterogeneous Network}
\newacronym{hh}{HH}{Hard Handover}
\newacronym{hol}{HOL}{Head-of-Line}
\newacronym{ia}{IA}{Initial Access}
\newacronym{imt}{IMT}{International Mobile Telecommunication}
\newacronym{iot}{IoT}{Internet of Things}
\newacronym{los}{LOS}{Line of Sight}
\newacronym{lte}{LTE}{Long Term Evolution}
\newacronym{m2m}{M2M}{Machine to Machine}
\newacronym{mac}{MAC}{Medium Access Control}
\newacronym{mc}{MC}{Multi-Connectivity}
\newacronym{mcs}{MCS}{Modulation and Coding Scheme}
\newacronym{mec}{MEC}{Mobile Edge Cloud}
\newacronym{mi}{MI}{Mutual Information}
\newacronym{mimo}{MIMO}{Multiple Input, Multiple Output}
\newacronym{mmwave}{mmWave}{millimeter wave}
\newacronym{mptcp}{MPTCP}{Multipath TCP}
\newacronym{mr}{MR}{Maximum Rate}
\newacronym{mss}{MSS}{Maximum Segment Size}
\newacronym{mtd}{MTD}{Machine-Type Device}
\newacronym{mtu}{MTU}{Maximum Transmission Unit}
\newacronym{nfv}{NFV}{Network Function Virtualization}
\newacronym{nlos}{NLOS}{Non Line of Sight}
\newacronym{nr}{NR}{New Radio}
\newacronym{ofdm}{OFDM}{Orthogonal Frequency Division Multiplexing}
\newacronym{pdcch}{PDCCH}{Physical Downlonk Control Channel}
\newacronym{pdcp}{PDCP}{Packet Data Convergence Protocol}
\newacronym{pdsch}{PDSCH}{Physical Downlink Shared Channel}
\newacronym{pdu}{PDU}{Packet Data Unit}
\newacronym{pf}{PF}{Proportional Fair}
\newacronym{pgw}{PGW}{Packet Gateway}
\newacronym{phy}{PHY}{Physical}
\newacronym{pbch}{PBCH}{Physical Broadcast Channel}
\newacronym[plural=\gls{mme}s,firstplural=Mobility Management Entities (MMEs)]{mme}{MME}{Mobility Management Entity}
\newacronym{prb}{PRB}{Physical Resource Block}
\newacronym{pss}{PSS}{Primary Synchronization Signal}
\newacronym{pucch}{PUCCH}{Physical Uplink Control Channel}
\newacronym{pusch}{PUSCH}{Physical Uplink Shared Channel}
\newacronym{rach}{RACH}{Random Access Channel}
\newacronym{ran}{RAN}{Radio Access Network}
\newacronym{red}{RED}{Random Early Detection}
\newacronym{rf}{RF}{Radio Frequency}
\newacronym{rlc}{RLC}{Radio Link Control}
\newacronym{rlf}{RLF}{Radio Link Failure}
\newacronym{rrc}{RRC}{Radio Resource Control}
\newacronym{rrm}{RRM}{Radio Resource Management}
\newacronym{rr}{RR}{Round Robin}
\newacronym{rs}{RS}{Remote Server}
\newacronym{rsrp}{RSRP}{Reference Signal Received Power}
\newacronym{rss}{RSS}{Received Signal Strength}
\newacronym{rtt}{RTT}{Round Trip Time}
\newacronym{rw}{RW}{Receive Window}
\newacronym{rx}{RX}{Receiver}
\newacronym{sa}{SA}{standalone}
\newacronym{sack}{SACK}{Selective Acknowledgment}
\newacronym{sap}{SAP}{Service Access Point}
\newacronym{sch}{SCH}{Secondary Cell Handover}
\newacronym{scoot}{SCOOT}{Split Cycle Offset Optimization Technique}
\newacronym{sdma}{SDMA}{Spatial Division Multiple Access}
\newacronym{sinr}{SINR}{Signal to Interference plus Noise Ratio}
\newacronym{sm}{SM}{Saturation Mode}
\newacronym{snr}{SNR}{Signal to Noise Ratio}
\newacronym{son}{SON}{Self-Organizing Network}
\newacronym{ss}{SS}{Synchronization Signal}
\newacronym{srs}{SRS}{Sounding Reference Signal}
\newacronym{sss}{SSS}{Secondary Synchronization Signal}
\newacronym{tb}{TB}{Transport Block}
\newacronym{tcp}{TCP}{Transmission Control Protocol}
\newacronym{tdd}{TDD}{Time Division Duplexing}
\newacronym{tdma}{TDMA}{Time Division Multiple Access}
\newacronym{tfl}{TfL}{Transport for London}
\newacronym{tm}{TM}{Transparent Mode}
\newacronym{trp}{TRP}{Transmitter Receiver Pair}
\newacronym{tti}{TTI}{Transmission Time Interval}
\newacronym{ttt}{TTT}{Time-to-Trigger}
\newacronym{tx}{TX}{Transmitter}
\newacronym{ue}{UE}{User Equipment}
\newacronym{ul}{UL}{Uplink}
\newacronym{uml}{UML}{Unified Modeling Language}
\newacronym{um}{UM}{Unacknowledged Mode}
\newacronym{utc}{UTC}{Urban Traffic Control}
\newacronym{vm}{VM}{Virtual Machine}
\newacronym{rsrq}{RSRQ}{Reference Signal Received Quality}
\newacronym{rssi}{RSSI}{Received Signal Strength Indicator}
\newacronym{crs}{CRS}{Cell Reference Signal}
\newacronym{comp}{CoMP}{Coordinated Multi-Point}
\newacronym{cran}{C-RAN}{Cloud \acrlong{ran}}
\newacronym{ca}{CA}{Carrier Aggregation}
\newacronym{cco}{CC}{Carrier Component}
\newacronym{nsa}{NSA}{Non Stand Alone}
\newacronym{embb}{eMBB}{Enhanced Mobility Broadband}
\newacronym{bsr}{BSR}{Buffer Status Report}
\newacronym{srb}{SRB}{Service Radio Bearer}
\newacronym{scm}{SCM}{Spatial Channel Model}

\usepackage{tikz}
\usepackage{pgfplots}
\pgfplotsset{compat=newest} 
\pgfplotsset{plot coordinates/math parser=false} 
\newlength\fheight
\newlength\fwidth
\usetikzlibrary{plotmarks,patterns,decorations.pathreplacing,backgrounds,calc,arrows,arrows.meta,spy,matrix}
\usepgfplotslibrary{patchplots,groupplots}
\usepackage{tikzscale}

\tikzstyle{startstop} = [rectangle, rounded corners, minimum width=2cm, minimum height=0.5cm,text centered, draw=black]
\tikzstyle{io} = [trapezium, trapezium left angle=70, trapezium right angle=110, minimum width=3cm, minimum height=1cm, text centered, draw=black]
\tikzstyle{process} = [rectangle, minimum width=2cm, minimum height=0.5cm, text centered, draw=black, align=center]
\tikzstyle{decision} = [ellipse, minimum width=2cm, minimum height=1cm, text centered, draw=black]
\tikzstyle{arrow} = [thick,<->,>=stealth]
\tikzstyle{line} = [thick,>=stealth]
\tikzstyle{darrow} = [thick,<->,>=stealth,dashed]
\tikzstyle{sarrow} = [thick,->,>=stealth]
\tikzstyle{larrow} = [line width=0.1mm,dashdotted,<->,>=stealth]

\definecolor{SchoolColor}{RGB}{0.71, 0, 0.106}
\definecolor{chaptergrey}{rgb}{0.61, 0, 0.09} 
\definecolor{midgrey}{rgb}{0.4, 0.4, 0.4}
\definecolor{chaptergreen}{rgb}{0.09, 0.612, 0}
\definecolor{chapterpurple}{rgb}{0.522, 0, 0.612}
\definecolor{chapterlightgreen}{rgb}{0, 0.612, 0.522}

\makeatletter
\def\grd@save@target#1{%
  \def\grd@target{#1}}
\def\grd@save@start#1{%
  \def\grd@start{#1}}
\tikzset{
  grid with coordinates/.style={
    to path={%
      \pgfextra{%
        \edef\grd@@target{(\tikztotarget)}%
        \tikz@scan@one@point\grd@save@target\grd@@target\relax
        \edef\grd@@start{(\tikztostart)}%
        \tikz@scan@one@point\grd@save@start\grd@@start\relax
        \draw[minor help lines] (\tikztostart) grid (\tikztotarget);
        \draw[major help lines] (\tikztostart) grid (\tikztotarget);
        \grd@start
        \pgfmathsetmacro{\grd@xa}{\the\pgf@x/1cm}
        \pgfmathsetmacro{\grd@ya}{\the\pgf@y/1cm}
        \grd@target
        \pgfmathsetmacro{\grd@xb}{\the\pgf@x/1cm}
        \pgfmathsetmacro{\grd@yb}{\the\pgf@y/1cm}
        \pgfmathsetmacro{\grd@xc}{\grd@xa + \pgfkeysvalueof{/tikz/grid with coordinates/major step x}}
        \pgfmathsetmacro{\grd@yc}{\grd@ya + \pgfkeysvalueof{/tikz/grid with coordinates/major step y}}
        \foreach \x in {\grd@xa,\grd@xc,...,\grd@xb}
        \node[anchor=north] at (\x,\grd@ya) {\pgfmathprintnumber{\x}};
        \foreach \y in {\grd@ya,\grd@yc,...,\grd@yb}
        \node[anchor=east] at (\grd@xa,\y) {\pgfmathprintnumber{\y}};
      }
    }
  },
  minor help lines/.style={
    help lines,
    gray,
    line cap =round,
    xstep=\pgfkeysvalueof{/tikz/grid with coordinates/minor step x},
    ystep=\pgfkeysvalueof{/tikz/grid with coordinates/minor step y}
  },
  major help lines/.style={
    help lines,
    line cap =round,
    line width=\pgfkeysvalueof{/tikz/grid with coordinates/major line width},
    xstep=\pgfkeysvalueof{/tikz/grid with coordinates/major step x},
    ystep=\pgfkeysvalueof{/tikz/grid with coordinates/major step y}
  },
  grid with coordinates/.cd,
  minor step x/.initial=.5,
  minor step y/.initial=.2,
  major step x/.initial=1,
  major step y/.initial=1,
  major line width/.initial=1pt,
}
\makeatother

\begin{document}


\flushbottom
\setlength{\parskip}{0ex plus0.1ex}

\title{Integration of Carrier Aggregation and Dual Connectivity\\for the ns-3 mmWave Module}




\author{\texorpdfstring{Tommaso Zugno, Michele Polese, Michele Zorzi\\
\small Department of Information Engineering, University of Padova, Padova, Italy \\
\small e-mail: \{zugnotom, polesemi, zorzi\}@dei.unipd.it}{}}

\copyrightyear{2018}
\setcopyright{none}
\acmConference[WNS3 2018]{the 2018 Workshop on ns-3}{June 2018}{NITK Surathkal, Mangalore, India}
\acmISBN{xxxxxxxxxxxx}
\acmPrice{xxxxx}
\acmDOI{xxxxxxxxxxxxxxxxxx}

\pagestyle{empty}

\begin{abstract}

Thanks to the wide availability of bandwidth, the millimeter wave (mmWave) frequencies will provide very high data rates to mobile users in next generation 5G cellular networks. However, mmWave links suffer from high isotropic pathloss and blockage from common materials, and are subject to an intermittent channel quality. Therefore, protocols and solutions at different layers in the cellular network and the TCP/IP protocol stack have been proposed and studied. A valuable tool for the end-to-end performance analysis of mmWave cellular networks is the ns-3 mmWave module, which already models in detail the channel, \gls{phy} and \gls{mac} layers, and extends the \gls{lte} stack for the higher layers. In this paper we present an implementation for the ns-3 mmWave module of multi connectivity techniques for 3GPP \gls{nr} at mmWave frequencies, namely \gls{ca} and \gls{dc}, and discuss how they can be integrated to increase the functionalities offered by the ns-3 mmWave module.

\end{abstract}


\maketitle

\begin{table*}[t]
	\centering	
	\caption{Multi connectivity solutions for mmWave cellular networks, at different layers and across a single or multiple \glspl{rat}.}
	\label{table:multiconn}
    \renewcommand{\arraystretch}{1.3}
    \setlength{\tabcolsep}{8pt} 
	\begin{tabularx}{0.95\textwidth}{>{\hsize=1\hsize}X>{\hsize=1.0\hsize}X>{\hsize=0.5\hsize}X>{\hsize=1.5\hsize}X}

	\toprule
	Multi connectivity technique & Relevant specifications for \gls{nr} at mmWave frequencies & Single or multi-\gls{rat} & Main features \\
	\midrule
	\acrlong{comp} & \gls{comp} is not included in 3GPP \gls{nr} specifications. Studied in~\cite{comp2016mmwave,comp2017mmwave,tesema2017multiconnectivity}. & Single & Increases the received \gls{snr} by combining multiple identical transmission \\

	\textbf{\acrlong{ca}} \newline\footnotesize (available in ns-3 mmWave) & 3GPP TS 38.300~\cite{38300}, TR 38.802~\cite{38802}. & Single & Increases the data rate or the diversity using multiple carriers with a common \acrshort{mac} layer \\	

	\textbf{\acrlong{dc}} \newline\footnotesize (available in ns-3 mmWave) & 3GPP TS 38.300~\cite{38300}, TS 37.340~\cite{37340} & Both & Uses different cells to increase the data rate or the reliability, and improve the mobility management. Used for \gls{lte}-\gls{nr} internetworking\\	

	\acrlong{mptcp} \newline\footnotesize (available in ns-3 with \gls{dce}~\cite{polese2017mptcp,tazaki2013dce}) & RFC 6824~\cite{rfc6824} (independent from 3GPP \gls{nr} specifications) & Multi & Combines multiple TCP subflows on different network interfaces to increase the throughput\\	

	Application layer solutions \newline\footnotesize (available in ns-3 with custom implementations~\cite{drago2018reliable}) & Independent from specification bodies. Studied in~\cite{drago2018reliable} & Multi & Use different \glspl{rat} to increase the diversity and improve the Quality of Experience\\ \bottomrule
		
	\end{tabularx}
\end{table*}
\section{Introduction}\label{sec:intro}
\glsresetall

\begin{picture}(0,0)(0,-500)
\put(0,0){
\put(0,0){\small This paper has been submitted to WNS3 2018. Copyright may be transferred without notice.}}
\end{picture}

Communication at mmWave frequencies will be one of the key features of the fifth generation of cellular networks (5G), given that the wide available bandwitdh at these frequencies~\cite{rappaportmillimeter} can potentially enable multi-gigabit-per-second data rates~\cite{BocHLMP:14,JerryPi:11} and satisfy the \acrlong{embb} 5G use case~\cite{shafi2017deployment}. 3GPP \gls{nr}, the 5G standard for cellular networks, will support the communication at frequencies up to 52.6 GHz~\cite{38802}, and the first trials have confirmed the potential for ultra-high achievable throughput~\cite{shafi2017deployment}. 

However, there are several challenges to be addressed for a successful deployment of mobile mmWave networks, mainly related to the harsh propagation environment at such high frequencies, and in the recent years there have been several efforts focused on solving these issues. The first is the high isotropic propagation loss, which increases with the square of the carrier frequency. This is addressed by using highly directional communications, that increase the link budget and hence the range at which the communication is still feasible, and are enabled by the fact that with a small mmWave wavelength it is possible to pack many antenna elements in small areas~\cite{samsBeamForm,ShSun14}. 
The second challenge is related to blockage, which prevents direct \gls{los} communication in the presence of obstacles, buildings and even the human body~\cite{RanRapE:14}. Nonetheless, as shown in~\cite{rappaportmillimeter}, in an urban environment with a rich scattering environment, it is possible to communicate also in \gls{nlos} using reflections, but with an \gls{snr} with is approximatively 30 dB smaller. This problem can be solved with network densification, i.e., the increment in the density of the deployment of mmWave base stations, with inter-site distance in the order of a few hundreds of meters to decrease the outage probability~\cite{RanRapE:14,JerryPi:11}. 

The characteristics of the communications at mmWave frequencies, however, also introduce challenges in the whole network stack. For example, the usage of directional transmissions requires new protocols for initial access and tracking at the \gls{mac} layer~\cite{shokri2015millimeter}, while the network densification and the sensitivity to blockage events call for fast network procedures to timely update the serving base station~\cite{poleseHo}. The performance of transport protocols is also affected by the intermittency of the mmWave channel, which causes the emergence of bufferbloat and low utilization of the available resources~\cite{polese2017tcp}.

The need for the design of cross-layer solution and the analysis of the performance of mmWave cellular networks in an end-to-end environment has motivated the introduction of a mmWave module for ns-3~\cite{ford2016framework}, which is publicly available and extensively described in~\cite{mezzavilla2017end}. It features the implementation of mmWave channel models (including the 3GPP model~\cite{zhang20173gpp}), and custom \gls{phy} and \gls{mac} layers with a dynamic frame structure that adapts to the large bandwidth available at mmWaves~\cite{Dutta:15}, and extends the higher layers of the \gls{lte} module implementation (e.g., by using queueing techniques in the \gls{rlc} buffers, or by modeling also the connection to the control elements of the core network). 
The module has been already used to study the performance of frame structures, schedulers, mobility management techniques and transport protocols in end-to-end mmWave cellular networks~\cite{mezzavilla2017end}.

In this paper, we describe the implementation of two multi connectivity techniques that are included in the latest 3GPP specifications for \gls{nr}~\cite{38300} and the internetworking between \gls{lte} and \gls{nr}~\cite{37340} which can be used to improve the connection reliability and/or the throughput. In particular, we will focus on \gls{ca} for mmWave frequencies, and describe how it can be integrated with multi-\gls{rat} \gls{dc} in a combined \gls{lte}-mmWave scenario. 
The inclusion of these features increases the realism and the capabilities of the ns-3 mmWave module, and enables the simulation of more complex scenarios with advanced multi-\gls{rat} solutions, agile spectrum management and higher throughput.

The remainder of the paper is organized as follows. In Sec.~\ref{sec:mc} we describe the main multi connectivity solutions for mmWave cellular networks, with a focus on \gls{ca} and \gls{dc} and the related 3GPP specifications. In Sec.~\ref{sec:dc} we present the implementation of \gls{ca}, and in Sec.~\ref{sec:ca} we discuss that of the \gls{dc} and the integration between the two. In Sec.~\ref{sec:examples} we report some examples and results, and we conclude the paper and provide insights on future works in Sec.~\ref{sec:concl}.

\section{Multi Connectivity for mmWave Cellular Networks}\label{sec:mc}
Given the harsh propagation environment at mmWave frequencies, and the probability of link disruption given by self-blockage or external obstacles, it is important to design and deploy mechanisms that provide diversity in the communication. Beside diversity in time, which can be achieved using retransmissions, and improves the overall end-to-end performance by hiding the channel losses to the higher layers~\cite{polese2017mptcp}, another important kind of diversity is introduced by multi connectivity over different base stations and/or with links at different frequencies (also called macro diversity)~\cite{RanRapE:14}. 
The multi connectivity for user and control planes can be implemented in multiple ways, which mainly differ for the layer at which the integration among the available links happens (i.e., \gls{phy}, \gls{mac} or higher layers) and the heterogeneity of the links (i.e., whether they belong to the same \gls{rat} or not), as shown in Table~\ref{table:multiconn}.

\paragraph*{Multi connectivity techniques for the same \gls{rat}}
At the physical layer, the same signals, transmitted from different synchronized access points of the same \gls{rat}, can be combined at the \gls{ue} side to increase the \gls{sinr} with the \gls{comp} technique. At mmWave frequencies, papers~\cite{comp2016mmwave,comp2017mmwave} analyze the gain in terms of coverage when using \gls{comp}, while the authors of~\cite{tesema2017multiconnectivity} study the throughput and \gls{rlf} performance in a \gls{cran} setup with \gls{comp}. 

At the \gls{mac} layer, instead, multi connectivity is usually achieved with \gls{ca}, which is already widely used in \gls{lte}-Advanced networks~\cite{pedersen2011ca,36300}. While with \gls{comp} the data transmitted by each base station is the same, in the same time and frequency resources, in order to increase the \gls{sinr}, with \gls{ca} different data streams can be transmitted in each link (also called \gls{cco}). Moreover, different \glspl{cco} can use different frequencies, and can be adapted to the channel independently (i.e., use different \glspl{mcs}, and/or retransmission processes), but are usually transmitted by the same base station. \gls{ca} increases the available datarate for the user, since it aggregates the spectrum across multiple bands, but can also be used for agile interference management~\cite{pedersen2011ca} and spectrum sharing with unlicensed bands with the LTE-U extension~\cite{zhang2015lteu}. \gls{ca} will be supported also by the 3GPP \gls{nr} standard, which also supports mmWave frequencies for the access, with a maximum of 16 \gls{cco}~\cite{38802,38300}. At mmWave frequencies, carrier aggregation techniques can be used to combine carriers with very different propagation properties (e.g., 28 and 73 GHz) or in licensed and unlicensed bands~\cite{khan2014carrier} in order to improve the reliability of transmission and/or increase the throughput~\cite{RanRapE:14}. However, to the best of our knowledge, there are no studies on the application of this technique to mmWaves.

Finally, another single-\gls{rat} multi-connectivity technique is \gls{dc}, introduced in 3GPP \gls{lte}-Advanced networks~\cite{36300} and extended in \gls{nr}~\cite{38300}. In this case, the user is connected to multiple base stations, with one cell acting as primary and the other as secondary. The integration happens at the \gls{pdcp} layer~\cite{38323}, which is located in the primary cell, and the lower layers (i.e., \gls{rlc}, \gls{mac} and \gls{phy}) of the different cells are independent. 

The main difference between \gls{dc} and \gls{ca} is in how the cellular network stack is set up for each bearer, i.e., for each end-to-end traffic flow. With \gls{dc}, each bearer is configured with a different and independent \gls{rlc} instance per cell, which forwards and receives data from a common \gls{pdcp} instance, while \gls{ca} has a single \gls{rlc} for each bearer (independently of the number of \glspl{cco}) and can support joint scheduling at the \gls{mac} layer across the different carriers. Therefore, while \gls{dc} can be deployed in different non-colocated cells (which could also use different multiplexing techniques and frame structures in the air interface), \gls{ca} is usually applied to carriers belonging to the same cell. However, \gls{ca} enables a tighter level of integration and balancing between the different links. With \gls{dc}, once the data is forwarded from the \gls{pdcp} to the primary or secondary \gls{rlc}, it will be transmitted in the selected cell and, if the channel quality of that link worsen, there is no possibility of changing the selected cell on the fly (the data would need to be forwarded from one \gls{rlc} to the other). With \gls{ca}, instead, there is a single \gls{rlc} layer for each bearer, thus, until the bearer data is actually scheduled on one of the available carriers, in principle it could be transmitted on any of them. Notice that \gls{dc} and \gls{ca} can be combined, i.e., a primary or a secondary cell can use multiple \glspl{cco} per user.


\paragraph*{Multi connectivity across different \glspl{rat}}
\acrlong{dc} plays an important role also in multi connectivity across different \glspl{rat}: for example, papers~\cite{dasilva,poleseHo} proposed \gls{dc} as a promising technique for the inter-networking between 4G (i.e., \gls{lte}) and 5G cellular networks, and the 3GPP recently announced the support of \gls{lte} and \gls{nr} integration with \gls{dc}~\cite{37340,38300}.
In particular, \gls{dc} between \gls{lte} and \gls{nr} is seen as a promising enabler of early \gls{nr} deployments, which would piggyback on the already deployed \gls{lte} core network (i.e., \gls{epc}), thus initially avoiding a costly deployment of the new 5G core network. Given that the integration is at the \gls{pdcp} layer, this option also allows a different design for the \gls{rlc}, \gls{mac} and physical layer in 5G \gls{nr} with respect to \gls{lte}. 

\begin{figure*}[t]
	\centering
	\includegraphics[width=.8\textwidth]{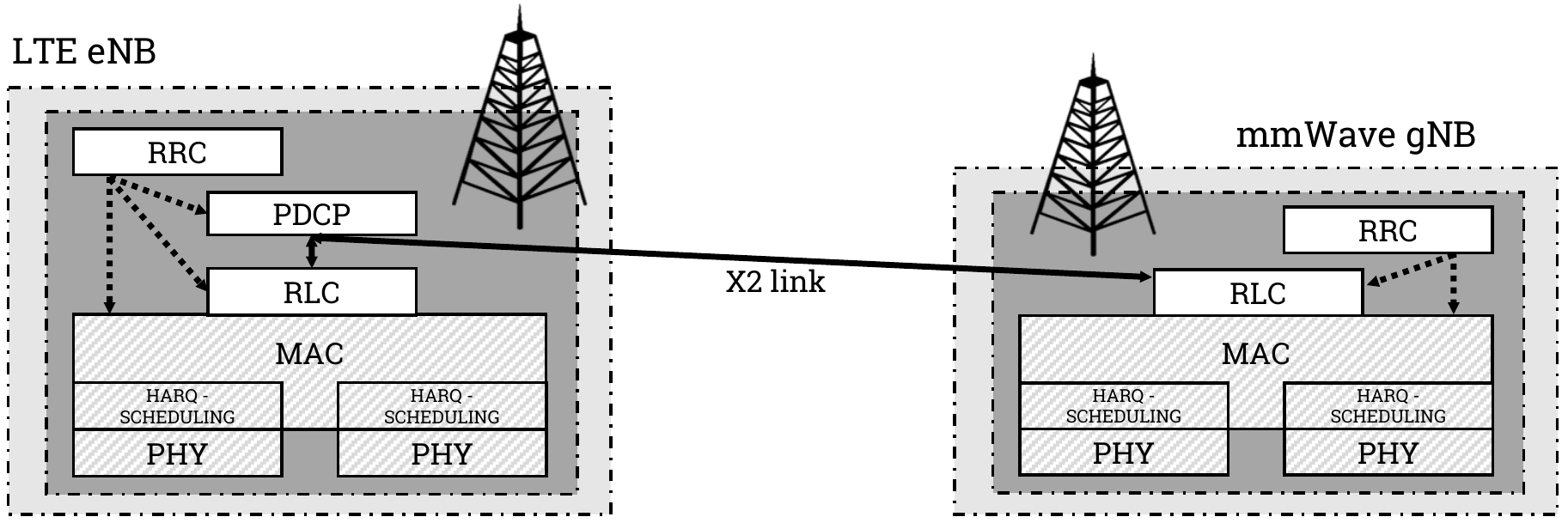}
	\caption{Protocol stack for the integration of multi-\gls{rat} dual connectivity and carrier aggregation at the \gls{ran} side. The layers in gray are those affected by the \gls{ca} implementation.}
	\label{fig:stack-enb}
\end{figure*}

\begin{figure}[t]
	\centering
	\includegraphics[width=\columnwidth]{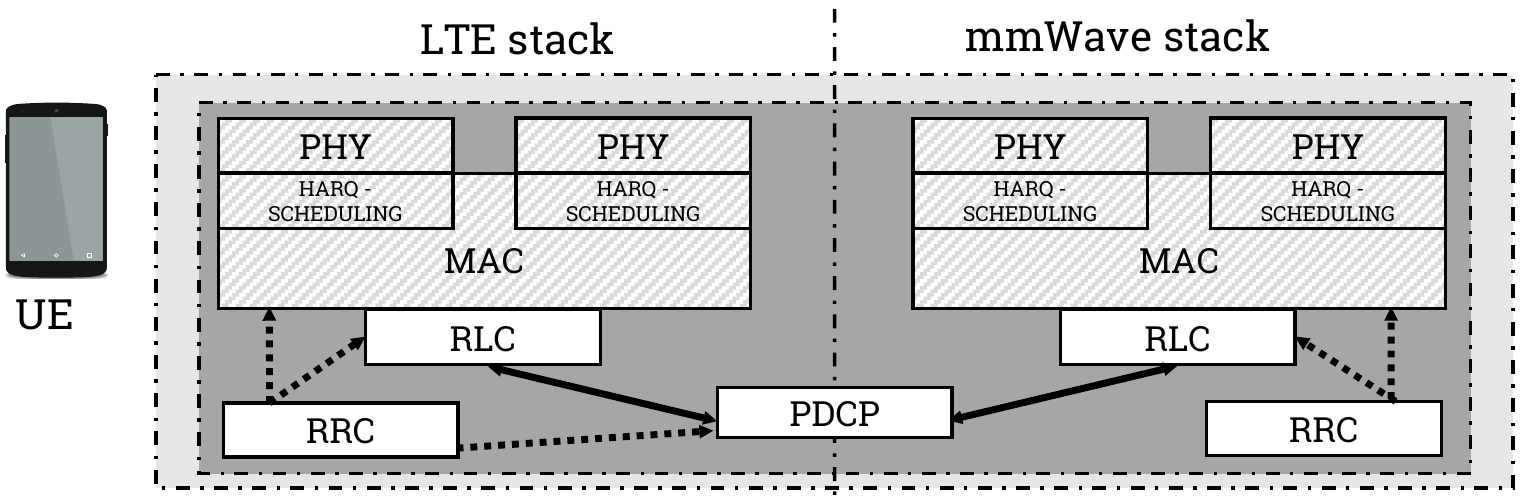}
	\caption{\gls{ue} protocol stack for the integration of multi-\gls{rat} dual connectivity and carrier aggregation. The layers in gray are those affected by the \gls{ca} implementation.}
	\label{fig:stack-ue}
\end{figure}

The performance of the combination of different \glspl{rat} at sub-6 GHz and mmWave frequencies with dual connectivity has been studied in~\cite{poleseHo,simutoolsPolese}, showing that it can improve both throughput stability and latency while reducing the control signaling and simplifying the mobility management. Moreover, multi connectivity across different \glspl{rat} can also improve the control plane reliability, for example by reducing the signaling overhead and the latency for the directional initial access needed at mmWave frequencies~\cite{giordaniMC2016}.

Finally, multi connectivity at the transport or application layer recently emerged as a possible enabler of simultaneous communication over different and completely independent \glspl{rat}, such as cellular networks managed by different operators and/or cellular and Wi-Fi networks. In particular, in~\cite{polese2017mptcp,saha2017poster} the performance of Multipath \gls{tcp}~\cite{rfc6824} has been studied over a combination of sub-6 GHz (LTE or Wi-Fi) and mmWave links, while in~\cite{drago2018reliable} multi connectivity on \gls{lte} and mmWave is used at the application layer to improve the quality of video streaming.

Given the importance of multi connectivity for mmWave networks, we believe that integrating enabling multi connectivity techniques in the mmWave module for ns-3 improves the realism and the validity of the performance evaluation of mmWave cellular networks. In the following sections we will present the multi connectivity solutions (i.e., multi \gls{rat} \acrlong{dc}, and \acrlong{ca}) available in ns-3 mmWave\footnote{Multipath \gls{tcp} can also be on top of different \glspl{rat}, e.g., mmWave and \gls{lte}, or Wi-Fi, using \gls{dce} as described in~\cite{mezzavilla2017end,polese2017mptcp}.}, as shown by the protocol stack represented for the \gls{ran} side in Fig.~\ref{fig:stack-enb} and the \gls{ue} in Fig.~\ref{fig:stack-ue}.

\section{Carrier Aggregation in ns-3 mmWave}\label{sec:ca}
The modeling of the \gls{ca} feature in the mmWave module for ns-3 follows the 3GPP specifications for \gls{nr}~\cite{38300}, and aligns the \gls{phy} and \gls{mac} design to the ns-3 \gls{lte} module implementation~\cite{lena}, for which the \gls{ca} capability was introduced in~\cite{bojovic2017towards}. In this section, we will describe the main characteristics of our implementation, and the differences with respect to \gls{ca} in the \gls{lte} module. 

As shown in Fig.~\ref{fig:stack-enb} and Fig.~\ref{fig:stack-ue}, the implementation for the data plane involves the lower layers of the protocols stack (i.e., \gls{mac} and \gls{phy}), i.e., it is transparent with respect to the functionalities offered by the \gls{rlc} and \gls{pdcp} layers. The control functionalities are performed by the \gls{rrc} layer, which is in charge of sharing the information for the carrier setup between the base station and the \gls{ue}. In particular, the base station broadcasts information on the primary \gls{cco}, and the \gls{ue} connects to it. Then, when it enters the RRC\_CONNECTED state, the base station \gls{rrc} can instruct the \gls{ue} to add and/or remove additional carriers with different parameters~\cite{38300}.

In our \gls{ca} model, and as generally done in the ns-3 mmWave module~\cite{mezzavilla2017end}, we inherit and extend the inter-layer interfaces of the \gls{lte} module (i.e., the \glspl{sap})~\cite{lena} and the classes that implement them, in order to increase the flexibility and account for different channel and propagation conditions for the different carriers, as well as possibly different numerologies, as specified in~\cite{38300}.

\begin{figure}[t]
	\centering	
	\includegraphics[width=.78\columnwidth]{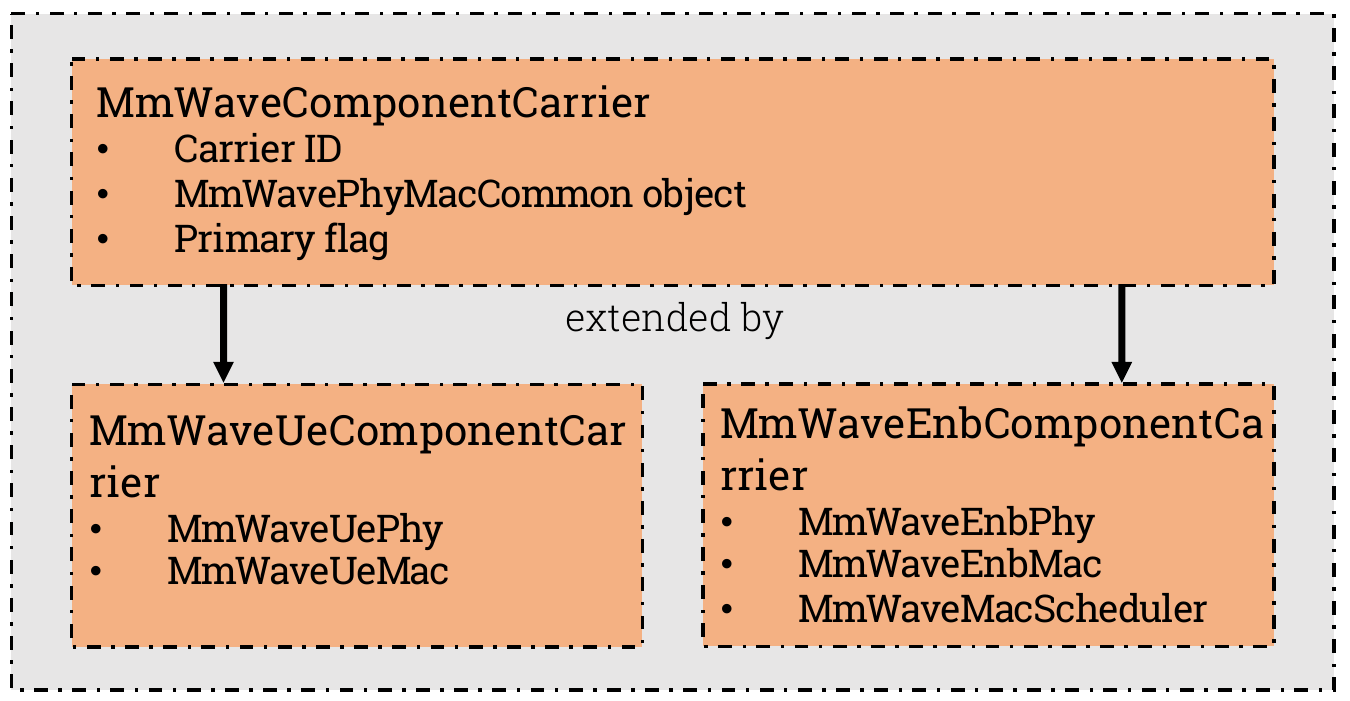}
	\caption{Information represented by instances of \texttt{MmWaveComponentCarrier} (and extensions).}
	\label{fig:cc}
\end{figure}

Similarly to the \gls{lte} implementation~\cite{bojovic2017towards}, the basic class of the \gls{ca} implementation is the \texttt{MmWaveComponentCarrier} class and its \texttt{MmWaveEnbComponentCarrier} and \texttt{MmWaveUeComponentCarrier} extensions. An instance of this class represents a single carrier, and contains pointers to the associated protocol stack layers and relevant configurations, as shown in Fig.~\ref{fig:cc}. In particular, in our implementation, a \texttt{MmWaveComponentCarrier} object contains a reference to a \texttt{MmWave\-PhyMacCommon} object, which is used to specify the numerology, frequency and bandwidth information for the carrier. The \texttt{MmWavePhy\-MacCommon} class was introduced in~\cite{MezzavillaNs3:15}, and prior to the \gls{ca} implementation, a single \texttt{MmWavePhyMacCommon} was created by \texttt{MmWaveHelper} during the configuration of the simulation. This object was shared by all the \gls{enb} and \gls{ue} \gls{phy} and \gls{mac} layer classes, as well as by the channel model classes, to provide access to a set of common parameters. With the \gls{ca} implementation, instead, an instance of \texttt{MmWavePhyMacCommon} is created for each possible carrier, and is associated to the unambiguous identifier of the carrier (i.e., carrier ID stored in the \texttt{m\_componentCarrierId} private variable of \texttt{MmWavePhyMacCommon}). Each of these objects is shared by all classes of the layers at the base station and \gls{ue} side that are related to the same carrier. The carrier-specific \texttt{MmWavePhyMac\-Common} instance then defines the carrier frequency (with the attribute \texttt{CenterFreq}), the bandwidth (for which it is possible to control the size of the resource blocks and their number) and the frame structure (i.e., the number of symbols per subframe, their duration, and the number of subframes per frame).

The different \texttt{MmWaveComponentCarrier} objects in the \glspl{ue} and base stations are managed by a single \gls{cco} manager, i.e., an object that implements respectively the \texttt{LteUeComponentCarrierManager} or the \texttt{LteEnbComponentCarrierManager} interfaces. The \gls{cco} manager, together with the \texttt{MmWaveUeMac} or \texttt{MmWaveEnbMac} classes, models the functionalities of the \gls{mac} layer as show in Fig.~\ref{fig:stack-enb} and Fig.~\ref{fig:stack-ue} for the mmWave protocol stack. In particular, at the base station side, it receives the \glspl{bsr} from the \gls{rlc} layers, and forwards them to the \texttt{MmWaveScheduler} instances following different policies according to the particular implementation of the \gls{cco} manager. The schedulers then allocate the available resources and generate \glspl{dci} for the different carriers. In the current implementation, the scheduling on the different carriers is independent, but we plan to extend it in order to model joint cross-carrier scheduling. The \gls{cco} manager at the \gls{ue} side is a simplified version of that of the base station, since it does not need to split the \glspl{bsr} between the carriers, but limits itself to forwarding them to the base station \gls{cco} manager. In particular, only the primary \gls{cco} is used for the reporting of the \glspl{bsr} and the exchange of control information\footnote{While implementing the \gls{ca} feature for the mmWave module, we discovered a bug that prevents the \gls{ca} implementation of the \gls{lte} module in ns-3.27 to use the resources allocated for the uplink. This bug is fixed in our implementation, and we proposed a patch also for \gls{lte}, see \url{https://www.nsnam.org/bugzilla/show_bug.cgi?id=2861}.}, since it is the only \gls{cco} in which the \glspl{srb} are set up.

In the mmWave \gls{ca} implementation, we provide different implementations of the \gls{cco} manager at the base station side. As for \gls{lte}, there is a \texttt{MmWaveNoOp\-MacComponentCarrierManager} which is used for single-carrier simulations, and a \texttt{MmWaveRrMacComponent\-CarrierManager}, which applies a round robin policy and splits the \glspl{bsr} equally across the carriers, with the result that they reach a similar throughput. In addition, we include also a bandwidth-aware \gls{cco} manager. It is likely that different carriers over different frequency bands will use different bandwidths, given that the higher the carrier frequency the larger the bandwidth that can be allocated to mobile network operators\footnote{For example, the International Telecommunication Union is considering the allocation to mobile operators of bands of approximatively 3 GHz in the 20--30 GHz spectrum, and of 10 GHz in the 60--80 GHz spectrum~\cite{itu2015resolution}.}. Therefore, a typical use case for \gls{ca} in the mmWave band would be the aggregation of a \gls{cco} at relatively low carrier frequency, with a smaller bandwidth, but with better propagation properties (i.e., lower pathloss), and other \glspl{cco} at much higher frequencies with larger bandwidths. In this case, a round robin \gls{cco} manager that evenly splits the packets to be transmitted across the different carriers would not yield an optimal performance, given the different data rates that can be supported by the \glspl{cc}. Therefore, the \gls{cco} manager implemented in the \texttt{MmWaveBaRrMacComponentCarrierManager} class is made aware of the bandwidth available to the different carriers during the simulation setup, and then, when it receives the \glspl{bsr} from the \gls{rlc} layer instances in the base station or the \gls{ue}, it divides the reports according to the bandwidth ratio across the carriers.

Another difference with respect to the \gls{lte} implementation is the usage of different channel model objects for the different carriers. In the mmWave module, indeed, the joint modeling of the propagation loss, the small and large scale fading and the beamforming has a fundamental importance for the accuracy of the simulation results. In our previous paper~\cite{zhang20173gpp} we introduced the implementation of the 3GPP channel model for frequencies above 6 GHz~\cite{38900}, which has features that depend on some carrier-specific parameters, such as the bandwidth and the carrier frequency. Therefore, we decided to use different \texttt{MmWave3gppChannel} objects\footnote{The \texttt{MmWave3gppChannel} class implements the \texttt{SpectrumPropagationLossModel} interface, and in particular the \texttt{DoCalcRxPowerSpectralDensity} method which applies fading and beamforming to the received power spectral density according to the 3GPP channel model and different beamforming techniques~\cite{zhang20173gpp}. Moreover, in this current iteration, we support only the 3GPP channel model, but we plan to extend the implementation to all the channel models available in ns-3 mmWave.} for each carrier, and use the \texttt{MmWavePhyMacCommon} of the carrier to set up the necessary parameters. Finally, we extended the \texttt{MmWaveSpectrumValueHelper} class in order to support the configuration (i.e., bandwidth, numerology and carrier frequency) of the different carriers.

\paragraph*{\gls{ca} configuration in ns-3 mmWave simulations}
Thanks to the adoption of a \texttt{MmWavePhyMacCommon} object per carrier, the user of the ns-3 mmWave module has a lot of flexibility in configuring the parameters of the simulation. We provide two comprehensive simulation examples in the \texttt{mmwave-ca-same-bandwidth.cc} and \texttt{mmwave-ca-diff-\-bandwidth.cc} files in the \texttt{examples} folder of the mmWave module. The first step in the simulation configuration is the initialization of a \texttt{MmWavePhyMacCommon} per \gls{cco}. The method \texttt{SetAttribute} can be used to set the relevant parameters for the carrier. Then, a map that associates the carrier ID to the \texttt{MmWaveComponentCarrier} is created, and passed as a parameter to the \texttt{MmWaveHelper} with the method \texttt{SetCcPhyParams}. The user then deploys the nodes, installs the relevant \texttt{NetDevices}, mobility models and applications as in a non-\gls{ca} simulation script. It is the \texttt{MmWaveHelper}, indeed, that transparently takes care of the initialization of the channel objects and the association to the correct carrier, and of the setup of the mmWave base stations and \glspl{ue} with the carriers information. 

\section{Dual Connectivity in ns-3 mmWave}\label{sec:dc}

The ns-3 mmWave module can also simulate \glspl{ue} that can connect to two different \glspl{rat} (i.e., \gls{lte} and mmWave base stations) at any given time. The modeling and the implementation of this functionality has been described in our previous works~\cite{simutoolsPolese,poleseHo,polese2016thesis}. For the sake of completeness, we recall here the main features, and describe the changes needed to support \gls{ca}. 

The \gls{dc} feature can be used to enhance the quality of the connection in two different ways, i.e., by increasing the reliability or the throughput. The reliability can be improved by using just the mmWave \gls{rat} and performing a seamless fallback to LTE  when all the mmWave base stations are in outage. The throughput, instead, is increased by actively transmitting data on both the links.
The implementation follows the 3GPP specifications for multi-\gls{rat} \gls{dc}~\cite{37340}, and involves the higher layers of the protocol stack shown in Fig.~\ref{fig:stack-enb} and Fig.~\ref{fig:stack-ue}, i.e., from the \gls{rlc} layer up. The core of the implementation is a new \texttt{NetDevice}, i.e., the \texttt{McUeNet\-Device}, which models a dual-stack (i.e., \gls{lte} and mmWave) \gls{ue} with a single \texttt{EpcUeNas} class as the interface between the higher layers and the cellular protocol stack in the data plane. 

\begin{figure}[t]
  \centering
  \begin{tikzpicture}[font=\sffamily\small, node distance=0.78cm, scale=.82, every node/.style={scale=.82}]
    \node [rectangle, minimum width=25em, minimum height=2em, draw=black, align=center] (pdcp) {\texttt{McEnbPdcp}};
    
    \node [rectangle, minimum width=12em, minimum height=2em, draw=black, anchor=east, align=center] at ($(pdcp.east) - (0, 3)$) (epcx2) {\texttt{EpcX2} on\\LTE eNB};

    \node [rectangle, minimum width=12em, minimum height=2em, draw=black, anchor=east, align=center, below of=epcx2, yshift=-2.5em] (epcx2rem) {\texttt{EpcX2} on\\mmWave eNB};

    \node [align=center, below of=epcx2] (label) {Point to Point link\\with latency, datarate};

    \node [rectangle, minimum width=12em, minimum height=2em, draw=black, anchor=west] at ($(pdcp.west) - (0, 3)$) (localRlc) {\texttt{LteRlc} or subclasses};

    \node [rectangle, minimum width=12em, minimum height=2em, draw=black, anchor=east] at ($(epcx2rem.east) - (0, 2.5)$) (remoteRlc) {\texttt{LteRlc} or subclasses};

    \draw [sarrow] ($(pdcp.south east) - (0.1, 0)$) -- node[anchor=east, align=right,yshift=-1.5em] {\texttt{EpcX2PdcpSapProvider}\\(\texttt{SendMcPdcpPdu})} ($(epcx2.north east) - (0.1, 0)$);

    \draw [sarrow] ($(epcx2.north east) - (3.7, 0)$) -- node[anchor=west, align=left,yshift=1.5em] {\texttt{EpcX2PdcpSapUser}\\(\texttt{ReceiveMcPdcpPdu})} ($(pdcp.south east) - (3.7, 0)$);

    \draw [sarrow] ($(pdcp.south west) + (0.1, 0)$) -- node[anchor=west, align=right, yshift=1.5em] {\texttt{LteRlcSapProvider}\\(\texttt{TransmitPdcpPdu})} ($(localRlc.north west) + (0.1, 0)$);

    \draw [sarrow] ($(localRlc.north west) + (3.7, 0)$) -- node[anchor=east, align=right,yshift=-1.5em] {\texttt{LteRlcSapUser}\\(\texttt{ReceivePdcpPdu})} ($(pdcp.south west) + (3.7, 0)$);

    \draw [sarrow] ($(epcx2rem.south east) - (0.1, 0)$) -- node[anchor=east, align=right, yshift=-1.5em] {\texttt{EpcX2RlcSapUser}\\(\texttt{SendMcRlcSdu})} ($(remoteRlc.north east) - (0.1, 0)$);

    \draw [sarrow] ($(remoteRlc.north east) - (3.7, 0)$) -- node[anchor=west, align=left,yshift=1.5em] {\texttt{EpcX2RlcSapProvider}\\(\texttt{ReceiveMcRlcSdu})} ($(epcx2rem.south east) - (3.7, 0)$);

    \draw [dashed] ($(epcx2.south west) + (0.1, 0)$) -- ($(epcx2rem.north west) + (0.1, 0)$);
    \draw [dashed] ($(epcx2.south east) - (0.1, 0)$) -- ($(epcx2rem.north east) - (0.1, 0)$);

  \end{tikzpicture}
  \caption{\gls{sap} interfaces for \gls{dc} between the \texttt{McEnbPdcp}, \texttt{LteRlc} (or subclasses) and \texttt{EpcX2} classes.}
  \label{fig:pdcp}
\end{figure}
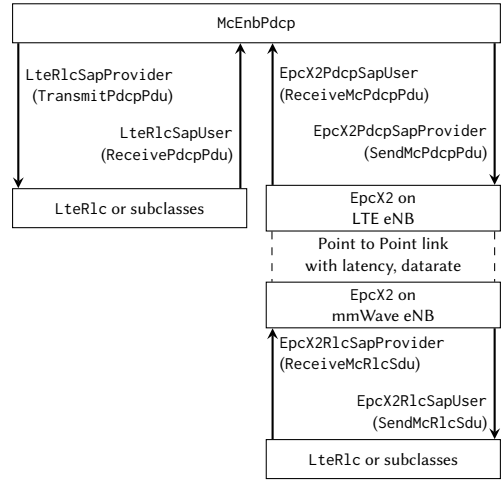

In order to simulate a \gls{nsa} deployment (i.e., a deployment of an \gls{lte} and of an \gls{nr} \gls{ran} with a common \gls{lte}-\gls{epc} core network), the \gls{lte} \gls{enb} acts as the primary cell, i.e., as the mobility anchor towards the core network, and the \gls{nr} \gls{gnb} at mmWave frequencies as the secondary cell. The \texttt{MmWaveHelper} class supports the installation of both \gls{lte} and mmWave base stations with the same \gls{epc}. The downlink packets for the \gls{ue} are routed from the \gls{pgw} to the \gls{lte} \gls{enb}, and the integration happens at the \gls{pdcp} layer: a single \gls{pdcp} instance is created for each bearer, and controls two instances of the \gls{rlc} layer, one in each cell. The X2 interface connects the different base stations, and is used to forward downlink \gls{pdcp} \glspl{pdu} from the primary to the secondary, and vice versa for the uplink. In order to support this functionality, we introduced two classes that extends the \texttt{LtePdcp} class, i.e., \texttt{McEnbPdcp} at the \gls{ran}  and \texttt{McUePdcp} at the \gls{ue} side. We also defined new \gls{sap} interfaces, between the \gls{pdcp}, the \gls{rlc} layer and X2, which are configured during the bearer setup process in the \gls{ran}, in order to enable the packet forwarding to and from the remote \gls{rlc}/\gls{pdcp} as shown in Fig.~\ref{fig:pdcp}.

The control plane was also extended in order to support \gls{dc}. Contrary to the intra-\gls{rat} \gls{dc} for \gls{lte}~\cite{36842}, in the latest \gls{lte}-\gls{nr} internetworking specifications~\cite{37340} and in our implementation~\cite{simutoolsPolese} both the \gls{lte} and \gls{nr} mmWave stacks feature a complete implementation of the \gls{rrc} layer. It performs control functionalities on the link such as initial access, collection and reporting of link measurements, bearer setup and management, mobility. The \gls{rrc} layer in the \gls{lte} stack manages both the local link and the setup of the dual connectivity with the selected mmWave cell. We extended the implementation of the \texttt{LteEnbRrc} and \texttt{LteUeRrc} classes to support the \gls{dc} functionalities, in particular for the new network procedures and the support of the split bearers. 

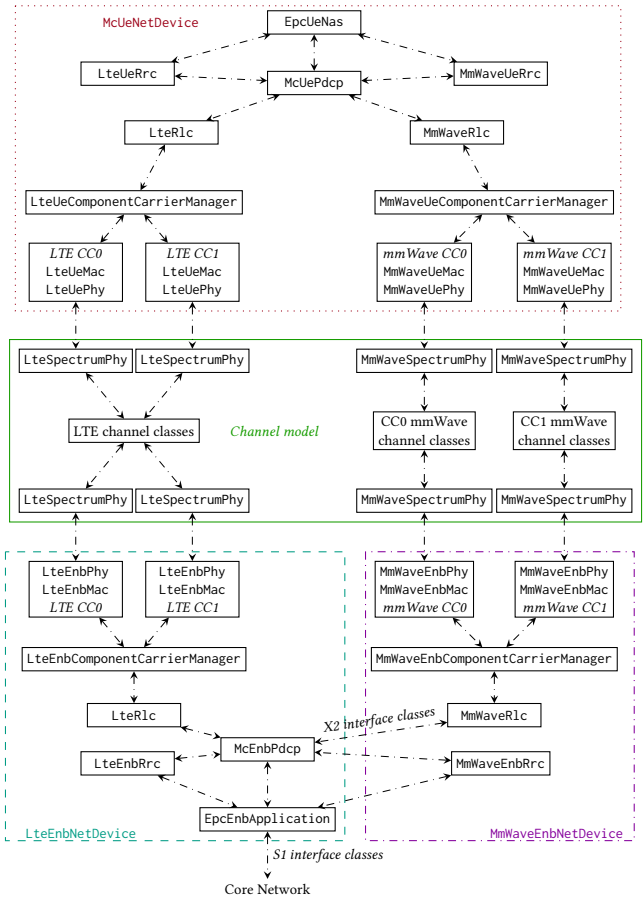
\begin{figure}[t]
\centering
\begin{tikzpicture}[node distance=1.5cm, scale=0.62, every node/.style={scale=0.62}]
  \node (epc) [process] {\texttt{EpcUeNas}};
  \node (rrc) [process, below of=epc, xshift=-4cm, yshift=0.4cm] {\texttt{LteUeRrc}};
  \node (mmrrc) [process, below of=epc, xshift=4cm, yshift=0.4cm] {\texttt{MmWaveUeRrc}};
  \node (pdcp) [process, below of=epc, yshift=0.2cm] {\texttt{McUePdcp}};
  \node (rlc) [process, below left of=pdcp, xshift=-2cm] {\texttt{LteRlc}};
  \node (mmrlc) [process, below right of=pdcp, xshift=2cm] {\texttt{MmWaveRlc}};

  \node (lteCc) [process, below of=rlc, yshift=0cm, xshift=-0.8cm] {\texttt{LteUeComponentCarrierManager}};

  \node (mmCc) [process, below of=mmrlc, yshift=0cm, xshift=0.8cm] {\texttt{MmWaveUeComponentCarrierManager}};

  \node (ltephy1) [process, below of=lteCc, xshift=-1.25cm] {\textit{LTE CC0}\\\texttt{LteUeMac} \\ \texttt{LteUePhy}};
  \node (ltephy1b) [process, below of=lteCc, xshift=1.25cm] {\textit{LTE CC1}\\\texttt{LteUeMac} \\ \texttt{LteUePhy}};
  \node (mmphy1) [process, below of=mmCc, xshift=-1.5cm] {\textit{mmWave CC0}\\\texttt{MmWaveUeMac} \\ \texttt{MmWaveUePhy}};
  \node (mmphy1b) [process, below of=mmCc, xshift=+1.5cm] {\textit{mmWave CC1}\\\texttt{MmWaveUeMac} \\ \texttt{MmWaveUePhy}};

  \draw[chaptergrey,dotted] ($(rrc.north west)+(-1.4,1.2)$) rectangle ($(mmphy1b.south east)+(0.2,-0.2)$);
  \node (legendMcUe) [left of=epc, xshift=-2cm]{\textcolor{chaptergrey}{\texttt{McUeNetDevice}}};

  \node (ltephy2) [process, below of=ltephy1, yshift=-0.4cm] {\texttt{LteSpectrumPhy}};
  \node (ltephy2b) [process, below of=ltephy1b, yshift=-0.4cm] {\texttt{LteSpectrumPhy}};
  \node (mmphy2) [process, below of=mmphy1, yshift=-0.4cm] {\texttt{MmWaveSpectrumPhy}};
  \node (mmphy2b) [process, below of=mmphy1b, yshift=-0.4cm] {\texttt{MmWaveSpectrumPhy}};

  \node (ltechannel) [process, below of=ltephy2, xshift=1.25cm] {LTE channel classes};
  \node (channel) [process, below of=mmphy2] {CC0 mmWave\\channel classes};
  \node (channelb) [process, below of=mmphy2b] {CC1 mmWave\\channel classes};

  \node (enbltephy2) [process, below of=ltechannel, xshift=-1.25cm] {\texttt{LteSpectrumPhy}};
  \node (enbltephy2b) [process, below of=ltechannel, xshift=1.25cm] {\texttt{LteSpectrumPhy}};
  \node (enbmmphy2) [process, below of=channel] {\texttt{MmWaveSpectrumPhy}};
  \node (enbmmphy2b) [process, below of=channelb] {\texttt{MmWaveSpectrumPhy}};

  \draw[chaptergreen,solid] ($(ltephy2.north west)+(-0.2,0.2)$) rectangle ($(enbmmphy2b.south east)+(+0.2,-0.2)$);
  \node (legendChannel) [right of=ltechannel, xshift=1.5cm] {\textit{\textcolor{chaptergreen}{Channel model}}};

  \node (lteEnbLl) [process, below of=enbltephy2, yshift=-0.4cm] {\texttt{LteEnbPhy} \\ \texttt{LteEnbMac}\\\textit{LTE CC0}};
  \node (lteEnbLlb) [process, below of=enbltephy2b, yshift=-0.4cm] {\texttt{LteEnbPhy} \\ \texttt{LteEnbMac}\\\textit{LTE CC1}};

  \node (mmEnbLl) [process, below of=enbmmphy2, yshift=-0.4cm] {\texttt{MmWaveEnbPhy} \\ \texttt{MmWaveEnbMac}\\\textit{mmWave CC0}};
  \node (mmEnbLlb) [process, below of=enbmmphy2b, yshift=-0.4cm] {\texttt{MmWaveEnbPhy} \\ \texttt{MmWaveEnbMac}\\\textit{mmWave CC1}};

  \node (lteEnbCc) [process, below of=lteEnbLl, xshift=1.25cm] {\texttt{LteEnbComponentCarrierManager}};

  \node (mmEnbCc) [process, below of=mmEnbLl, xshift=1.5cm] {\texttt{MmWaveEnbComponentCarrierManager}};

  \node (lteEnbRlc) [process, below of=lteEnbCc, yshift=0.3cm] {\texttt{LteRlc}};
  \node (mmEnbRlc) [process, below of=mmEnbCc, yshift=0.3cm] {\texttt{MmWaveRlc}};
  \node (enbPdcp) [process, below of=pdcp, yshift=-12.8cm, xshift=-1cm] {\texttt{McEnbPdcp}};
  \node (enbrrc) [process, below of=rrc, yshift=-13.3cm] {\texttt{LteEnbRrc}};
  \node (enbmmrrc) [process, below of=mmrrc, yshift=-13.3cm] {\texttt{MmWaveEnbRrc}};
  \node (epcEnb) [process, below of=enbPdcp] {\texttt{EpcEnbApplication}};
  \node (cn) [below of=epcEnb] {Core Network};

  \draw[chapterpurple,dashdotted] ($(mmEnbLl.north west)+(-0.2,0.2)$) rectangle ($(enbmmrrc.south east)+(1.8,-1.4)$);
  \node (legendChannel) [below of=enbmmrrc, xshift=1.2cm] {\textcolor{chapterpurple}{\texttt{MmWaveEnbNetDevice}}};

  \draw[chapterlightgreen,dashed] ($(lteEnbLl.north west)+(-0.5,0.2)$) rectangle ($(epcEnb.south east)+(0.2,-0.2)$);
  \node (legendChannel) [below of=enbrrc, xshift=-1cm] {\textcolor{chapterlightgreen}{\texttt{LteEnbNetDevice}}};

  \draw[larrow] (epc) -- (rrc);
  \draw[larrow] (epc) -- (mmrrc);
  \draw[larrow] (epc) -- (pdcp);
  \draw[larrow] (mmrrc) --  (pdcp);
  \draw[larrow] (rrc) --  (pdcp);
  \draw[larrow] (pdcp) -- (rlc);
  \draw[larrow] (pdcp) -- (mmrlc);
  \draw[larrow] (rlc) -- (lteCc);
  \draw[larrow] (mmrlc) -- (mmCc);
  \draw[larrow] (mmCc) -- (mmphy1);
  \draw[larrow] (mmCc) -- (mmphy1b);
  \draw[larrow] (lteCc) -- (ltephy1);
  \draw[larrow] (lteCc) -- (ltephy1b);

  \draw[larrow] (ltephy1) -- (ltephy2);
  \draw[larrow] (ltephy1b) -- (ltephy2b);
  \draw[larrow] (mmphy1) -- (mmphy2);
  \draw[larrow] (mmphy1b) -- (mmphy2b);

  \draw[larrow] (mmphy2) -- (channel);
  \draw[larrow] (mmphy2b) -- (channelb);
  \draw[larrow] (ltephy2b) -- (ltechannel);
  \draw[larrow] (ltephy2) -- (ltechannel);

  \draw[larrow] (channel) -- (enbmmphy2);
  \draw[larrow] (channelb) -- (enbmmphy2b);
  \draw[larrow] (ltechannel) -- (enbltephy2);
  \draw[larrow] (ltechannel) -- (enbltephy2b);

  \draw[larrow] (enbltephy2) -- (lteEnbLl);
  \draw[larrow] (enbltephy2b) -- (lteEnbLlb);
  \draw[larrow] (enbmmphy2) -- (mmEnbLl);
  \draw[larrow] (enbmmphy2b) -- (mmEnbLlb);

  \draw[larrow] (lteEnbLl) -- (lteEnbCc);
  \draw[larrow] (lteEnbLlb) -- (lteEnbCc);
  \draw[larrow] (mmEnbLl) -- (mmEnbCc);
  \draw[larrow] (mmEnbLlb) -- (mmEnbCc);

  \draw[larrow] (mmEnbCc) -- (mmEnbRlc);
  \draw[larrow] (lteEnbCc) -- (lteEnbRlc);

  \draw[larrow] (mmEnbRlc) -- node[sloped, anchor=center, above] {\textit{X2 interface classes}} (enbPdcp);
  \draw[larrow] (lteEnbRlc) -- (enbPdcp);
  \draw[larrow] (enbmmrrc) --  (enbPdcp);
  \draw[larrow] (enbrrc) --  (enbPdcp);
  \draw[larrow] (enbmmrrc) --  (epcEnb);
  \draw[larrow] (enbrrc) --  (epcEnb);
  \draw[larrow] (enbPdcp) -- (epcEnb);

  \draw[larrow] (epcEnb) -- node[anchor=west] {\textit{S1 interface classes}} (cn);


\end{tikzpicture}
\caption{Simplified UML of a dual-connected device, an LTE eNB and a MmWave eNB that also support carrier aggregation. We only report the main classes of the \gls{dc}-\gls{ca} integration implementation, i.e., the \gls{sap} interfaces are omitted.}
\label{fig:mcdevice}
\end{figure}

\begin{figure*}[t]
\centering
\begin{subfigure}[t]{0.49\textwidth}
	\centering	
	\setlength\fwidth{0.9\columnwidth}
	\setlength\fheight{0.4\columnwidth}
%
%
\definecolor{mycolor1}{rgb}{0.90471,0.19176,0.19882}%
\definecolor{mycolor2}{rgb}{0.29412,0.54471,0.74941}%
\definecolor{mycolor3}{rgb}{0.37176,0.71765,0.36118}%
\definecolor{mycolor4}{rgb}{1.00000,0.54824,0.10000}%
\begin{tikzpicture}
\pgfplotsset{every tick label/.append style={font=\scriptsize}}

\begin{axis}[%
width=0.951\fwidth,
height=\fheight,
at={(0\fwidth,0\fheight)},
scale only axis,
xmin=45,
xmax=155,
xtick=data,
xlabel style={font=\footnotesize\color{white!15!black}},
xlabel={Distance [m]},
ymin=0,
ymax=1.5,
ylabel style={font=\footnotesize\color{white!15!black}},
ylabel={${S}_{\rm RLC}$ [Gbit/s]},
axis background/.style={fill=white},
axis x line*=bottom,
axis y line*=left,
xmajorgrids,
ymajorgrids,
xminorgrids,
yminorgrids,
legend style={at={(0.01,0.96)},anchor=south west,font=\footnotesize,legend cell align=left, align=left, draw=white!15!black}
]
\addplot [color=mycolor1, dashed, mark=o, mark options={solid, mycolor1}]
 plot [error bars/.cd, y dir = both, y explicit]
 table[row sep=crcr, y error plus index=2, y error minus index=3]{%
50	1.36953973976	0.0261049591080033	0.0261049591080033\\
100	0.73214940896	0.0851283920113788	0.0851283920113788\\
150	0.3231877512	0.0213979796183118	0.0213979796183118\\
};
\addlegendentry{2 \gls{cc}, $B=500$~MHz, $f_0 = 39.75$~GHz, $f_1=40.25$~GHz}

\addplot [color=mycolor2, mark=x, mark options={solid, mycolor2}]
 plot [error bars/.cd, y dir = both, y explicit]
 table[row sep=crcr, y error plus index=2, y error minus index=3]{%
50	1.36995032928	0.0191523207068568	0.0191523207068568\\
100	0.72293139776	0.0839431389739547	0.0839431389739547\\
150	0.31254080024	0.0170876996475383	0.0170876996475383\\
};
\addlegendentry{2 \gls{cc}, $B=500$~MHz, $f_0 = 39.75$~GHz, $f_1=40.25$~GHz in blockage}

\addplot [color=mycolor3, dashed, mark=asterisk, mark options={solid, mycolor3}]
 plot [error bars/.cd, y dir = both, y explicit]
 table[row sep=crcr, y error plus index=2, y error minus index=3]{%
50	1.21628614856	0.0480334842576769	0.0480334842576769\\
100	0.53484166632	0.0371936762032177	0.0371936762032177\\
150	0.20517690512	0.0178195890557594	0.0178195890557594\\
};
\addlegendentry{1 \gls{cc}, $B=1$~GHz, $f_0 = 40$~GHz}

\addplot [color=mycolor4, mark=diamond, mark options={solid, mycolor4}]
 plot [error bars/.cd, y dir = both, y explicit]
 table[row sep=crcr, y error plus index=2, y error minus index=3]{%
50	1.17241253832	0.048989622268887	0.048989622268887\\
100	0.47726166688	0.0448836992277999	0.0448836992277999\\
150	0.180492806398238	0.0210382528485651	0.0210382528485651\\
};
\addlegendentry{1 \gls{cc}, $B=1$~GHz, $f_0 = 40$~GHz in blockage}

\end{axis}
\end{tikzpicture}%
	\caption{Contiguous allocation: 2 carriers with a bandwidth of 500 MHz each, at 39.75 and 40.25 GHz, the second with and without blockage, or 1 carrier with a bandwidth of 1 GHz at 40 GHz with and without blockage.}
	\label{fig:scenario1-contiguous}
\end{subfigure}\hfill%
\begin{subfigure}[t]{0.49\textwidth}
	\centering	
	\setlength\fwidth{0.9\columnwidth}
	\setlength\fheight{0.4\columnwidth}
	o
%
%
\definecolor{mycolor1}{rgb}{0.90471,0.19176,0.19882}%
\definecolor{mycolor2}{rgb}{0.29412,0.54471,0.74941}%
\definecolor{mycolor3}{rgb}{0.37176,0.71765,0.36118}%
\definecolor{mycolor4}{rgb}{1.00000,0.54824,0.10000}%
\begin{tikzpicture}
\pgfplotsset{every tick label/.append style={font=\scriptsize}}

\begin{axis}[%
width=0.951\fwidth,
height=\fheight,
at={(0\fwidth,0\fheight)},
scale only axis,
xmin=45,
xmax=155,
xtick=data,
xlabel style={font=\footnotesize\color{white!15!black}},
xlabel={Distance [m]},
ymin=0,
ymax=1.5,
ylabel style={font=\footnotesize\color{white!15!black}},
ylabel={${S}_{\rm RLC}$ [Gbit/s]},
axis background/.style={fill=white},
axis x line*=bottom,
axis y line*=left,
xmajorgrids,
ymajorgrids,
xminorgrids,
yminorgrids,
legend style={at={(0.01,0.96)},anchor=south west,font=\footnotesize,legend cell align=left, align=left, draw=white!15!black}
]
\addplot [color=mycolor1, dashed, mark=o, mark options={solid, mycolor1}]
 plot [error bars/.cd, y dir = both, y explicit]
 table[row sep=crcr, y error plus index=2, y error minus index=3]{%
50	1.26541047888	0.0258261257305578	0.0258261257305578\\
100	0.62950690848	0.0481251128680599	0.0481251128680599\\
150	0.28520073544	0.0180181904182918	0.0180181904182918\\
};
\addlegendentry{2 \gls{cc}, $B=500$~MHz, $f_0 = 32.5$~GHz, $f_1=73$~GHz}

\addplot [color=mycolor2, mark=x, mark options={solid, mycolor2}]
 plot [error bars/.cd, y dir = both, y explicit]
 table[row sep=crcr, y error plus index=2, y error minus index=3]{%
50	1.2526066044	0.0224287082433159	0.0224287082433159\\
100	0.57731830168	0.0216522282207731	0.0216522282207731\\
150	0.25984229392	0.0167971601116371	0.0167971601116371\\
};
\addlegendentry{2 \gls{cc}, $B=500$~MHz, $f_0 = 32.5$~GHz, $f_1=73$~GHz in blockage}

\addplot [color=mycolor3, dashed, mark=asterisk, mark options={solid, mycolor3}]
 plot [error bars/.cd, y dir = both, y explicit]
 table[row sep=crcr, y error plus index=2, y error minus index=3]{%
50	0.8712241452	0.0494726623703017	0.0494726623703017\\
100	0.24633815176	0.024595246152955	0.024595246152955\\
150	0.0738821413149278	0.0116505551652124	0.0116505551652124\\
};
\addlegendentry{1 \gls{cc}, $B=1$~GHz, $f_0 = 73$~GHz}

\addplot [color=mycolor4, mark=diamond, mark options={solid, mycolor4}]
 plot [error bars/.cd, y dir = both, y explicit]
 table[row sep=crcr, y error plus index=2, y error minus index=3]{%
50	0.8234520252	0.048949193856924	0.048949193856924\\
100	0.22012997576	0.0262597456618064	0.0262597456618064\\
150	0.0602758051411411	0.0129376313456516	0.0129376313456516\\
};
\addlegendentry{1 \gls{cc}, $B=1$~GHz, $f_0 = 73$~GHz in blockage}

\end{axis}
\end{tikzpicture}%
	\caption{Non contiguous allocation: 2 carriers with a bandwidth of 500 MHz each, at 32.5 and 73 GHz, the second with and without blockage, or 1 carrier with a bandwidth of 1 GHz at 73 GHz with and without blockage.}
	\label{fig:scenario1-different}
\end{subfigure}
\caption{Throughput at the \gls{rlc} layer for different configurations of the carrier aggregation in the \texttt{mmwave-ca-same-bandwidth.cc} example.}
\label{fig:scenario1}
\end{figure*}
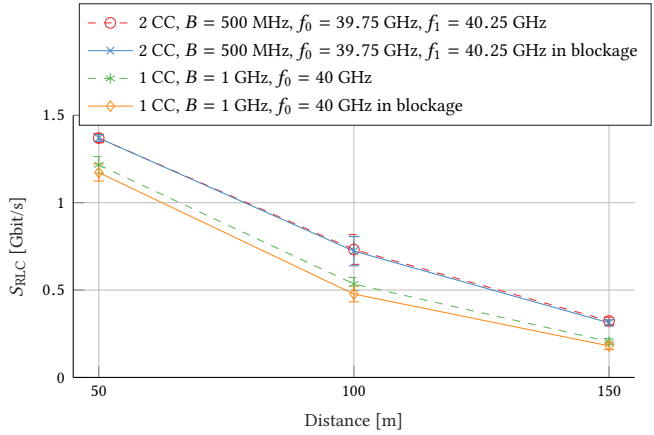
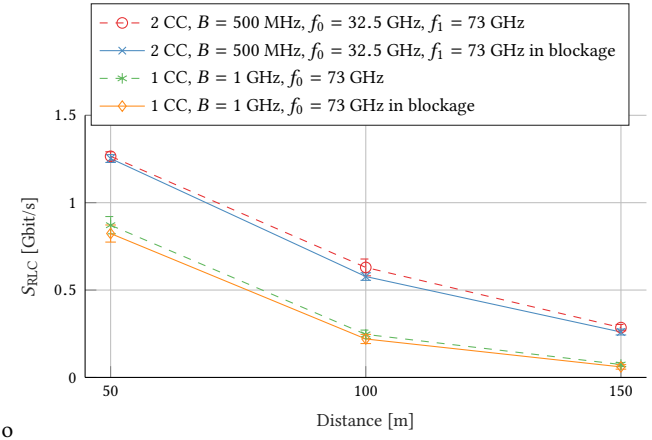

The ns-3 mmWave \gls{dc} also includes a fast handover procedure for the secondary cell (i.e., the one at mmWave), extensively described in~\cite{poleseHo}, that avoids the interaction with the core network during secondary cell updates and improves the performance of the mmWave \gls{ran} with mobile \glspl{ue} by reducing the handover interruption time and the latency during handover events. Moreover, during the handover procedure, we support two different \gls{rlc} buffer management policies. If \gls{rlc} \gls{am} is used, the handover is lossless, i.e., the \glspl{pdu} in all the \gls{rlc} buffers are forwarded from the source to the target cell. Instead, when \gls{um} is adopted, the handover is seamless, i.e., the source forwards to the targer cell only the \gls{rlc} \glspl{pdu} which have not yet been transmitted~\cite{sesia2011lte,llho}.

With respect to the implementation described in~\cite{simutoolsPolese}, we extended the \texttt{McUeNetDevice} class in order to support carrier aggregation. Fig.~\ref{fig:mcdevice} shows a simplified UML diagram of the integration of the \gls{dc} and \gls{ca} implementation for an \texttt{McUeNetDevice} and an \gls{lte} and a mmWave base stations. It can be seen that the mmWave and the \gls{lte} \gls{ca} implementations are used respectively in the \texttt{MmWaveEnbNetDevice} and \texttt{LteEnbNetDevice} classes, while they coexist in the \texttt{McUeNetDevice}. The example in Fig.~\ref{fig:mcdevice} shows two \glspl{cco} per \gls{rat}, but it is actually possible to configure independently the number of \glspl{cco} in the \gls{lte} and mmWave \glspl{rat}. Then, given that different \glspl{rrc} are in control of the \gls{lte} and mmWave links, it is possible to set the carriers after the \gls{ue} has attached to either of the two \glspl{rat}, using \gls{rrc} connection reconfiguration messages.

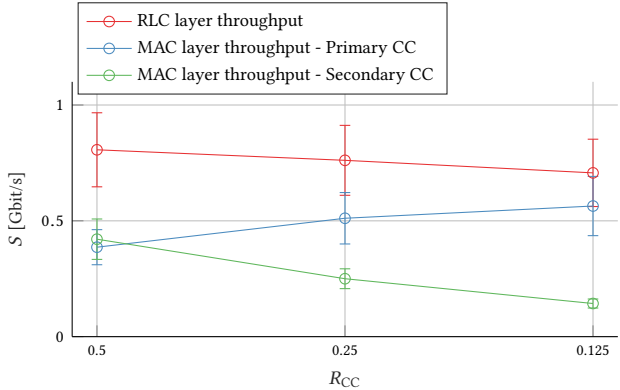
\begin{figure}[t]
	\centering	
	\setlength\fwidth{0.9\columnwidth}
	\setlength\fheight{0.4\columnwidth}
%
%
\definecolor{mycolor1}{rgb}{0.90471,0.19176,0.19882}%
\definecolor{mycolor2}{rgb}{0.29412,0.54471,0.74941}%
\definecolor{mycolor3}{rgb}{0.37176,0.71765,0.36118}%
\begin{tikzpicture}
\pgfplotsset{every tick label/.append style={font=\scriptsize}}

\begin{axis}[%
width=0.951\fwidth,
height=\fheight,
at={(0\fwidth,0\fheight)},
scale only axis,
xmin=0.9,
xmax=3.1,
xtick=data,
xticklabels={{0.5},{0.25},{0.125}},
xlabel style={font=\footnotesize\color{white!15!black}},
xlabel={$R_{\rm CC}$},
ymin=0,
ymax=1.1,
ylabel style={font=\footnotesize\color{white!15!black}},
ylabel={$S$ [Gbit/s]},
axis background/.style={fill=white},
axis x line*=bottom,
axis y line*=left,
xmajorgrids,
ymajorgrids,
xminorgrids,
yminorgrids,
legend style={at={(0.01,0.95)},anchor=south west,font=\footnotesize,legend cell align=left, align=left, draw=white!15!black}
]
\addplot [color=mycolor1, mark=o, mark options={solid, mycolor1}]
 plot [error bars/.cd, y dir = both, y explicit]
 table[row sep=crcr, y error plus index=2, y error minus index=3]{%
1	0.80667710336	0.159820796881591	0.159820796881591\\
2	0.76118826688	0.1509565827977	0.1509565827977\\
3	0.70704258376	0.145351248899872	0.145351248899872\\
};
\addlegendentry{RLC layer throughput}

\addplot [color=mycolor2, mark=o, mark options={solid, mycolor2}]
 plot [error bars/.cd, y dir = both, y explicit]
 table[row sep=crcr, y error plus index=2, y error minus index=3]{%
1	0.386184716	0.0757500755734094	0.0757500755734094\\
2	0.511083326	0.11096258514084	0.11096258514084\\
3	0.56392592536	0.127899837113083	0.127899837113083\\
};
\addlegendentry{MAC layer throughput - Primary \gls{cc}}

\addplot [color=mycolor3, mark=o, mark options={solid, mycolor3}]
 plot [error bars/.cd, y dir = both, y explicit]
 table[row sep=crcr, y error plus index=2, y error minus index=3]{%
1	0.42074540024	0.0872162619556569	0.0872162619556569\\
2	0.25008519864	0.0428091985233099	0.0428091985233099\\
3	0.14307614064	0.0194239091790635	0.0194239091790635\\
};
\addlegendentry{MAC layer throughput - Secondary \gls{cc}}

\end{axis}
\end{tikzpicture}%
	\caption{Comparison (in the \texttt{mmwave-ca-diff-bandwidth.cc} example) among different bandwidth splits for the two \glspl{cco}, i.e., for $R_{\rm CC} = B_{\rm CC1} / B_{\rm CC0} \in [0.5, 0.25, 0.125]$, with a constant total bandwidth $B = B_{\rm CC0} + B_{\rm CC1} = 1$~GHz in the $39.5-40.5$ GHz spectrum.}
	\label{fig:scenario2}
\end{figure}

\section{Examples}\label{sec:examples}

In this section we describe two examples related to the application of \gls{ca} to mmWave frequencies and report some relevant results that illustrate the flexibility of the \gls{ca} implementation in configuring carrier-specific parameters and of the \gls{cco} manager. They can be found in the files \texttt{mmwave-ca-same-bandwidth.cc} and \texttt{mmwave-ca-diff-\-bandwidth.cc}.

In the following examples, we always consider a total bandwidth of 1 GHz. However, we compare different scenarios in which one \gls{cco} (with a fraction of the bandwidth) or the whole 1 GHz bandwidth is affected by additional blockage. 
The 3GPP channel model~\cite{38900,zhang20173gpp}, indeed, models different kinds of blockage phenomena. The main distinction is between the \gls{nlos} and \gls{los} conditions, which differ because the main cluster of rays\footnote{The 3GPP 3D \gls{scm} considers the received signal as a combination of different \textit{clusters}, composed of multiple \textit{rays}. Each cluster has its own angle of arrival and departure, delay with respect to the first cluster, and phase, and the total power is given by the aggregation of the power in the different clusters~\cite{38900}.} is blocked or not, respectively. However, it is possible to model additional blockage events on the other clusters. This feature represents the attenuation that can be caused by (i) the human body of the user holding the device (self-blockage), or (ii) other external obstacles (non-self-blockage). The blockage is randomly applied in certain angular directions, and thus on the clusters whose angle of arrival or departure belongs to those regions. The per-cluster attenuation specified in the 3GPP model and in our implementation is of 30 dB~\cite{38900}, even though some recent papers proposed measurement-based models with a smaller attenuation (i.e., 15 dB)~\cite{raghavan2018statistical}. 
The additional blockage can be set using the \texttt{Blockage} attribute of the \texttt{MmWave3gppChannel} class, and there is the possibility of selectively setting it carrier by carrier with the method \texttt{SetBlockageMap} of the \texttt{MmWaveHelper} class. In our performance evaluation, we consider that, if multiple carriers are used with different antenna arrays in the mobile device, then the user may just block one of them with his/her hand or body, while, if the whole bandwidth is allocated to a single carrier, then the link is either completely blocked or not. 

In the first example, we consider a single user in a \gls{nlos} condition with respect to the serving mmWave base station. They exchange data both in downlink and uplink, using a full buffer condition at the \gls{rlc} layer that saturates the capacity of the link (i.e., the \gls{rlc} instance belongs to the class \texttt{LteRlcSm}). The user is placed at a 2D distance $d \in [50, 100, 150]$~m from the base station, with an urban macro fading condition~\cite{38900}. In this example, we consider one or two \gls{cco} using the same total amount of bandwidth. Therefore, if a single carrier is selected, it is configured with a bandwidth $B=1$~GHz, whereas, if two carriers are set up, each of them will use a bandwidth $B=500$~MHz. We also compare two \gls{cco} deployment strategies. The first is a contiguous allocation in the $37 - 40.5$ GHz band, around a carrier frequency of 40 GHz. The second, instead, is a non-contiguous deployment with a \gls{cco} at 32.5 GHz, in the $31.8 - 33.4$ GHz band, and the other at 73 GHz, in the $66-76$ GHz band~\cite{itu2015resolution}. The same number of antenna elements for both configurations is used at the base station (64) and at the \gls{ue} (16).

The results are shown in Fig.~\ref{fig:scenario1-contiguous} for the contiguous allocation, and in Fig.~\ref{fig:scenario1-different} for the other one. We only report the downlink throughput $S_{\rm RLC}$, since the results in uplink are similar given the frame structure adopted. It can be immediately seen that the usage of \gls{ca} improves the \gls{rlc} throughput in both cases. If \gls{ca} is not used, and the whole 1 GHz bandwidth is controlled by a single scheduler, then it is not possible to properly adapt to different channel conditions that may be present in different chunks of the allocated bandwidth. The metric that is used to allocate the \gls{mcs}, and consequently the transmission opportunities, is indeed based on the average \gls{sinr}, and does not take into account frequency selective fading. With \gls{ca}, instead, it is possible to assign different \glspl{mcs} to each \gls{cco}, that maximize the throughput while preventing packet loss, and different retransmission processes~\cite{38300}, so that each \gls{cco} can optimally adjust to the different channel conditions.
This can be observed also when comparing the contiguous allocation case in Fig.~\ref{fig:scenario1-contiguous} with the non-contiguous one in Fig.~\ref{fig:scenario1-different}. 
In the first configuration, the spectrum band of the scenarios with and without \gls{ca} is the same (from 39.5 to 40.5 GHz), therefore the pathloss and fading parameters are similar. In the non-contiguous one, instead, one of the \glspl{cco} or the only carrier if \gls{ca} is not used are at 73 GHz. It can be seen that the \gls{ca} manages to make up for the throughput loss given by the higher pathloss at 73 GHz, and that the gap between the scenarios with and without \gls{ca} is larger in the non-contiguous configuration than in the contiguous. Finally, especially in the contiguous deployment scenario, the \gls{ca} throughput is similar when the secondary \gls{cco} is in blockage or not, corroborating the performance gain that can be achieved with a more agile resource allocation and channel adaptation mechanism. 

The second example, instead, uses the bandwidth-aware scheduler, and compares the performance of different ratios $R_{\rm CC} = B_{\rm CC1} / B_{\rm CC0} \in [0.5, 0.25, 0.125]$ between the bandwidth allocated to the different carriers. The \gls{ca} deployment is in a contiguous spectrum band around the 40 GHz carrier. The user is randomly placed at a distance $d \in \mathcal{U}[0, 150]$~m and moves in the scenario with a random walk mobility model. Fig.~\ref{fig:scenario2} reports the downlink \gls{mac}-layer throughput of each carrier and the downlink \gls{rlc} throughput. It can be seen that the performance worsen as the ratio $R_{\rm CC}$ decreases, and one of the two carriers occupies a much larger bandwidth than the other. For the secondary \gls{cco}, whose bandwidth decreases with $R_{\rm CC}$, the ratio between the throughput and the allocated bandwidth remains constant, while, for the primary, whose bandwidth increases with $R_{\rm CC}$, the same ratio decreases.
A configuration with a small $R_{\rm CC}$ is indeed similar to a configuration without \gls{ca} for the primary \gls{cco}, and does not provide the same channel adaptation capabilities as a configuration with a more even split of the bandwidth between the two \glspl{cc}. 

\section{Conclusions}\label{sec:concl}
In this paper, we presented the first implementation of carrier aggregation for the ns-3 mmWave module, and the integration of \gls{ca} with the \gls{lte}-\gls{nr} dual connectivity feature. Multi connectivity is an important feature in mmWave cellular networks, since it helps increase the reliability of the mmWave link by providing macro diversity (i.e., the possibility of using multiple mmWave links with different frequencies and spatial characteristics) and a ready fallback to legacy networks at sub-6 GHz features. Therefore, modeling both \gls{ca} and \gls{dc} in the ns-3 mmWave module is an important contribution to the module, that makes it possible to simulate more complex, advanced and realistic scenarios\footnote{The code of the \gls{dc}-\gls{ca} implementation can be found at \url{https://github.com/nyuwireless-unipd/ns3-mmwave/tree/ca-dc-integration}}. 

After an overview on the multi connectivity options for mmWave in the literature and in the 3GPP or IETF specifications, we described the implementation of \gls{ca}, focusing on the flexibility of the configuration of the parameters in the different carriers and on the implementation of a bandwidth-aware carrier manager. Then, we illustrated the \gls{dc} implementation, with additional details on the integration with \gls{ca}. Finally, we provided some examples and preliminary results for \gls{ca} at mmWave frequencies, showing how \gls{ca} improves the throughput of the network even if the same total bandwidth is considered, given the higher efficiency in performing the per-carrier scheduling and the macro diversity.

As future work, we plan to investigate additional \gls{cco} manager policies, which could benefit from \gls{phy}-\gls{mac} cross-layer approaches, with additional channel information considered in the allocation of resources to the different carriers. Moreover, we will implement joint-carrier schedulers, to increase the efficiency of \gls{ca}, and complete our implementation with a test suite. 

\bibliographystyle{ACM-Reference-Format.bst}
\bibliography{bibl.bib}

\end{document}